\documentclass[10pt,onecolumn,prb,aps,notitlepage]{revtex4}

\usepackage{tikz}
\usepackage{amssymb}
\usepackage{psfrag}
\usepackage{textcomp}
\usepackage{pdfsync}
\usepackage{amsmath}
\usepackage{amsfonts}
\usepackage{graphicx,bm,epstopdf}
\usepackage{subfigure}
\usepackage{comment}
\usepackage{verbatim}
\usepackage{color}
\usepackage{upgreek}
\usepackage{tikz}
\usepackage{natbib,hyperref}
\usepackage{threeparttable}
\usepackage{ifpdf}

\begin{document}

\title{Tailoring Magnetic Skyrmions by Geometric Confinement of Magnetic Structures}
\author{Steven S.-L. Zhang}
\email{shulei.zhang@anl.gov}
\affiliation{Material Science Division, Argonne National Laboratory, Lemont, Illinois 60439, USA}
\affiliation{Department of Physics and Astronomy, University of Missouri, Columbia, Missouri 65211, USA}
\author{C. Phatak}
\affiliation{Material Science Division, Argonne National Laboratory, Lemont, Illinois 60439, USA}
\author{A. K. Petford-Long}
\affiliation{Material Science Division, Argonne National Laboratory, Lemont, Illinois 60439, USA}
\affiliation{Department of Materials Science and Engineering, Northwestern University, Evanston, Illinois 60208, USA}
\author{O. G. Heinonen}
\affiliation{Material Science Division, Argonne National Laboratory, Lemont, Illinois 60439, USA}
\affiliation{Northwestern-Argonne Institute of Science and Technology, 2145 Sheridan Road, Evanston, Illinois 60208, USA}

\date{\today }

\begin{abstract}
Nanoscale magnetic skyrmions have interesting static and transport properties that make them candidates for future spintronic devices. 
Control and manipulation of the size and behavior of skyrmions is thus of crucial importance. Using a Ginzburg-Landau approach, we show theoretically that skyrmions and skyrmion lattices can be stabilized by a spatial modulation of the uniaxial magnetic anisotropy in a thin film of centro-symmetric ferromagnet. Remarkably, the skyrmion size is determined by the ratio of
the exchange length and the period of the spatial modulation of the anisotropy, at variance with conventional skyrmions stabilized by dipolar and Dzyaloshinskii--Moriya interactions (DMIs).
\end{abstract}

\maketitle

\section{Introduction}
Two dimensional
magnetic skyrmions are nanoscale spin textures that are
topologically protected: the spin structure of an
individual skyrmion is associated with an integer winding number which
cannot be continuously changed into another integer number without
overcoming a finite energy barrier
\cite{Fert17NatComm-Sk-review,Muhlbauer09Sci_skyrmion,Nagaosa13Nat.Nano_Skyrmion,Hoffmann17RMP-sk,Miltat16PRB_Sk-TI-barrier}%
. The creation, annihilation and transport of magnetic skyrmions strongly
rely on their topological properties~\cite{Okubo12PRL_sk-fluc-ex,Heinze11NatPhys_sk-4spin,Jiang15Sci-bubble-sk,heinonen16PRB_sk-SH,Woo16NatMater_sk-thinFilm,Boulle16NatNano_sk-thinFilm,cd16SciRep_skyrm_CoPt}, which make them promising candidates
as spin information carriers in future spintronic devices.

Multiple formation mechanisms of magnetic skyrmions have been identified~%
\cite%
{Lin73APL_bubble-garnet,Takao83JMMM_bubble-FM-film,slon79book-bubble,Muhlbauer09Sci_skyrmion,Okubo12PRL_sk-fluc-ex,Heinze11NatPhys_sk-4spin,Jiang15Sci-bubble-sk,heinonen16PRB_sk-SH}%
. Most commonly, stable skyrmions are found in bulk chiral magnets such as
MnSi~\cite%
{Muhlbauer09Sci_skyrmion,Lebech95JMMM_SkX-A-phase,Pfleiderer12PRB_MnSi-SkX}
and other B20 transition metal alloys~\cite%
{Grigoriev09PRL_DM-FeCoSi,Uchida08PRB_Skrm-FeGe,Yu&Tokura11NatMater_skrm-FeGe,Wilhelm11PRL_Helimag-FeGe,ShibataK.13NatNano_skrm-helimag}%
. In these systems, strong spin-orbit coupling conspires with broken bulk
inversion symmetry to give rise to the Dzyaloshinskii--Moriya interactions
(DMIs)~\cite{DZYALOSHINSKY58JPCS_DMintxn,Moriya60PR_DM-intxn} that favor
canted spin structure and thus can stabilize skyrmions with definite
chirality.
DMI can also be induced in a nonchiral transition metal thin film
in contact with a heavy metal layer~\cite%
{Emori13Nat.Mater_thin-film-interfacial-DMI,Ryu&Parkin13NatNano_thinFilmDMI}. This
gives rise to structural inversion-symmetry breaking and strong interfacial spin
orbit interaction, which can support the formation of skyrmions.
Even in the
absence of DMI, long range dipolar interactions alone may stabilize skyrmions, or magnetic bubbles,
as well, but the size of this type of skyrmion ($\sim 0.1$ to $1$ $\mu m$) ~%
\cite{slon79book-bubble,Nagaosa13Nat.Nano_Skyrmion}, is usually larger than
that stabilized by DMI as it scales with the ratio
of the exchange coupling to the dipolar interaction. This type of skyrmion will not, in the absence of DMI, have a distinct chirality, and both chiralities are degenerate in energy.

Crystallization of skyrmions occurs when the inter-skyrmion distance is
sufficiently reduced so that the repulsive skyrmion-skyrmion interaction leads to a packing in a hexagonal lattice.~\cite%
{McMillan75_sk-CDW,Hubert94_sk-Interaction,Leonov16NJP_sk-intxn}
A skyrmion crystal phase (SkX) has been observed both in thin films of chiral
magnets~\cite{Lebech95JMMM_SkX-A-phase,Bauer12PRB_MnSi-SkX} and magnetic
multilayer with perpendicular anisotropy~\cite{Fullerton17PRB_dipole-sk}. In
chiral magnets with very low Curie temperatures, the SkX phase is stabilized at
temperatures well below room temperature and requires an external magnetic field
(of the order of $1~T$~\cite{Nagaosa13Nat.Nano_Skyrmion,Fert17NatComm-Sk-review}) perpendicular to the film plane. In case of magnetic thin films as well, a magnetic field perpendicular to the film plane is required for the strip domain ground state to evolve into a (chiral) bubble lattice~\cite{Jiang15Sci-bubble-sk,Woo16NatMater_sk-thinFilm,Fert16Nat.Nano_add-DMI,Boulle16NatNano_sk-thinFilm,cd16SciRep_skyrm_CoPt}.

Better control of the physical properties of magnetic skyrmions, such as
their stability, size, chirality, etc., is not only of fundamental
interest but also crucial for the application of skyrmions in
spintronics. Some recent research has focused on this control. Small
individual skyrmions (with diameters smaller than 100~nm) were stabilized at
room temperature by additive interfacial DMIs~\cite{DZYALOSHINSKY58JPCS_DMintxn,Moriya60PR_DM-intxn} in Pt$\mid $Co$\mid $Ir
multilayers~\cite{Fert16Nat.Nano_add-DMI}.
Nucleation of magnetic skyrmions with a wide range of sizes and ellipticities was recently observed in a wedge-shaped FeGe nanostripe~\cite{cmJin17NatComm_FeGeWedge-skym}. Montoya \textit{et.~al.} experimentally demonstrated~\cite{Montoya17PRB_dipole-skrm}
that the stability and size of skyrmions originating from dipolar
interaction can be controlled by tuning the magnetic properties such as the
magnitude of the perpendicular uniaxial and shape anisotropies of Fe$\mid$%
Gd multilayers.

In addition to chiral magnets and transition-metal
multilayers, stable skyrmions may also be hosted
in centrosymmetric systems including
various multiferroic
materials~\cite{Seki12sci_skrm-multiferroic,Yu&Nagaosa12helicity-reversal-Sk,Langner&Tokura14PRL_skrmCuOSeO,white14PRL_skyrm-CuOSeO,Phatak16NanoLett_skyrm-multiferro}. An inherent advantage of multiferroic materials is that they
are characterized by more than one order parameter, which may impart unique
features to the skyrmions. For example, helicity reversal inside skyrmions was
observed in Sc-doped hexagonal barium ferrite~\cite{Yu&Nagaosa12helicity-reversal-Sk}, a multiferroic with tunable
magnetic anisotropy. More recently, a room temperature SkX state was observed
in thin films of the centro-symmetric material Ni$_{2}$MnGa~\cite{Phatak16NanoLett_skyrm-multiferro}
in the absence of DMI. Surprisingly, in that work all skyrmions in each realization of a
SkX state had the same chirality, but different realizations may exhibit
different chiralities, in contrast with the SkXs stabilized via DMI and
dipolar interactions. The degeneracy of the different chiralities opens up the possibility of controlling and altering the chirality of such SkXs, which may enable new applications. Phatak and coworkers associated the formation of this
type of SkX with the geometric confinement of magnetic structures by narrow
twin variants\cite{Phatak16NanoLett_skyrm-multiferro} with alternating in-plane and out-of-plane uniaxial magnetic anisotropy.

\begin{figure}[tph]
\centering
\includegraphics[trim={0.2cm 0.2cm 0.2cm 0.2cm},clip=true,
width=0.45\textwidth]{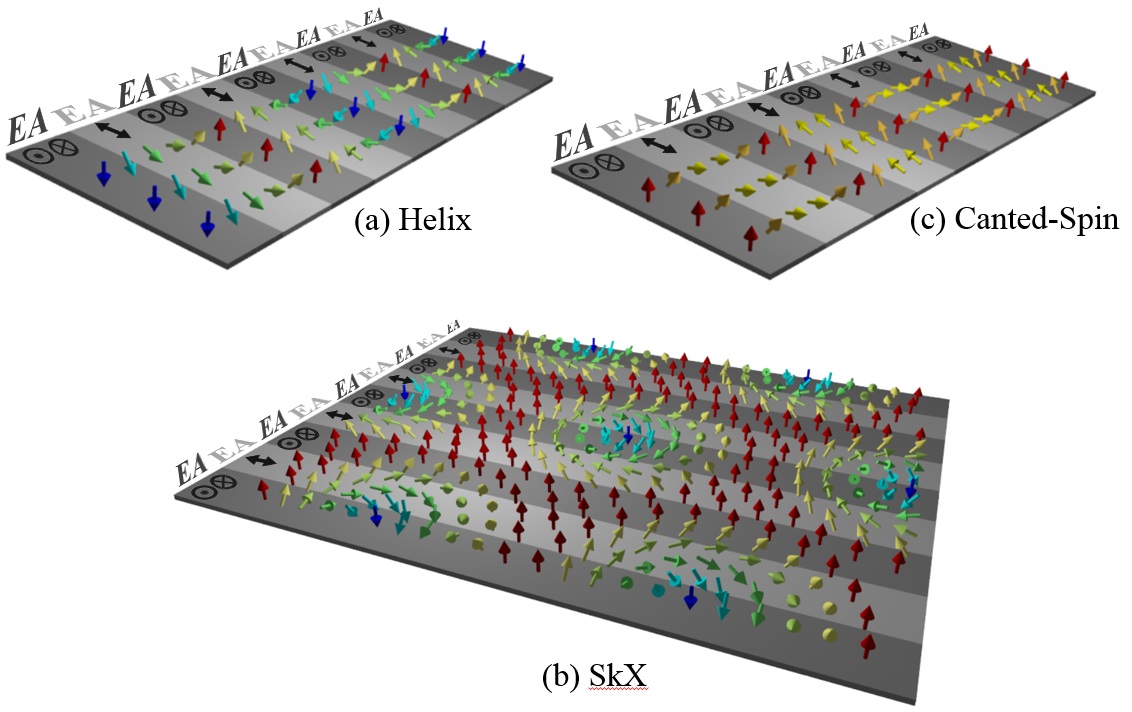}
\caption{Schematics of three typical magnetic states in a thin film where the anisotropy exhibits periodic spatial variation of easy axis (EA) from in-plane (denoted by $\leftrightarrow$) to out-of-plane (denoted by $\odot \otimes$).}
\label{fig:schematics}
\end{figure}

Inspired by these experimental works, we investigate theoretically the
energy landscape of various magnetic states in
a ferromagnetic thin film that arises from the competition between
exchange, shape, and modulated uniaxial anisotropy with a goal of understanding the phase diagram and under what conditions a Skx can be stabilized by these competing energy terms. We generalize the Ginzburg-Landau (GL) theory~\cite{Garel82PRB_Ginzberg-Landau} for
an Ising ferromagnet to take into account the general three
dimensional magnetization in a ferromagnetic thin film.
In particular, we will
consider a novel uniaxial anisotropy with periodic in-plane to out-of-plane spatial variation of the easy
axis, which can be realized in a
multiferroic material such as Ni$_{2}$MnGa~\cite{Phatak16NanoLett_skyrm-multiferro}. Intuitively, this form of
anisotropy favors a helical ground state as well as a canted-spin state with
an antiferromagnetic
arrangement in consecutive in-plane anisotropy twins, as shown schematically in
Fig.~\ref{fig:schematics}; we shall show that the SkX can form during the transition between these two dominant
magnetic phases. We will discuss the stability of the SkX phase as a
function of temperature and external magnetic field. Furthermore, we derive an explicit expression for the anisotropy energy density of the SkX.
This allows us to determine the size-dependence of the skyrmions on the ratio of the exchange length and the period of the spatial variation of the anisotropy, which is a hallmark of this unconventional SkX.

\section{Ginzburg-Landau model}
We assume that the film lies in the $x-y$ plane, and that the film thickness
is sufficiently thin so that the magnetization density $\mathbf{M}$ is uniform along the
$z$-direction and hence is a function only of $x$ and $y$, i.e., $\mathbf{M=M%
}\left( x,y\right) $. In the GL theory, the spatial average of the total magnetic free
energy ${\cal F}_{tot}$ of a ferromagnetic thin film of area $S$ may be written as
\begin{align}
\mathcal{F}_{tot} =S^{-1}\int &\mathrm{d}^{2}\mathbf{x}\left[ A_{ex}\left(
\nabla \mathbf{m}\right) ^{2}+t\mathbf{m}^{2}+u\left( \mathbf{m}^{2}\right)
^{2}\right. \notag   \\
&\left. -\mu_0\mathbf{m\cdot H}M_{0}+K_{d}m_{z}^{2}+f_{a}\left( \mathbf{m}%
\right) \right] \,, \label{Eq:f_tot}
\end{align}
where $\mathbf{m=M}/M_{0}$ with $|{\mathbf M}|=M(T)$ the local magnetization density at temperature $T$ and $M_0=|{\mathbf M}|(T\to0)$, $t$ and $u$ are
GL parameters that are in general functions of temperature
and external magnetic field, $\mu _{0}$ is the magnetic permeability, and $%
K_{d}m_z^2=\frac{1}{2}\mu _{0}M_{0}^{2}m_z^2$ denotes the demagnetizing energy density
in the thin film approximation~\cite{Harte68JAP_demag-thinFilm}. We shall draw particular attention
to the anisotropy energy density $f_{a}\left( \mathbf{m}\right) $. The
uniaxial magnetic anisotropy with spatially varying easy axis may be modeled
as
\begin{equation}
f_{a}\left[ \mathbf{m}\left( \mathbf{x}\right) \right] =-K_u\left[ \varkappa_{+} \left(
x\right) m_{z}^{2}+\varkappa_{-}\left( x\right) m_{y}^{2} \right]\,,
\label{Eq: f_a[m(r)]}
\end{equation}%
where $\varkappa_{\pm}(x)\equiv \frac{1}{2}\left[ \left\vert \cos \left( \frac{\pi x}{w_{t}}\right) \right\vert \pm  %
\cos \left( \frac{\pi x}{%
w_{t}}\right) \right] $ with $w_{t}$ is
the twin width, and $K_{u}\left( >0\right) $ characterizes the magnitude of the anisotropy.

A generalized spatial profile of the magnetization of a skyrmion lattice can
be approximated as a superposition of three spin helices~\cite%
{Muhlbauer09Sci_skyrmion,Nagaosa13Nat.Nano_Skyrmion}
\begin{align}
\mathbf{m}\left( \mathbf{r}\right)  =&\left[ m_{0}+m_{i,\perp
}\cos \left( \mathbf{k}_{i}\cdot \mathbf{r}\right) \right] \,\hat{\mathbf{z}}%
\notag \\
&+m_{i,\parallel }\left( \hat{\mathbf{z}}\times \mathbf{\hat{k%
}}_{i}\right) \sin \left( \mathbf{k}_{i}\cdot \mathbf{r}\right)\,,  \label{Eq: m(r)-profile}
\end{align}%
where $m_{0}$ is the uniform magnetization induced by the external magnetic
field perpendicular to the film plane, $m_{i,\perp }$ and $%
m_{i,\parallel }$ are the out-of-plane and in-plane components of the
magnetization, and we use Einstein's summation convention over repeated indices. 
The wave vectors of the three helices are all in the plane of
the layer and form an angle of $%
120^{\circ }$ with each other; explicitly, we choose $\mathbf{k}_{1}=q\hat{%
\mathbf{x}}$, $\mathbf{k}_{2}=\left( -\frac{1}{2}\hat{\mathbf{x}}+\frac{%
\sqrt{3}}{2}\hat{\mathbf{y}}\right) q$ and $\mathbf{k}_{3}=\left( -\frac{1}{2%
}\hat{\mathbf{x}}-\frac{\sqrt{3}}{2}\hat{\mathbf{y}}\right) q$, and $\mathbf{%
\hat{k}}_{i}=\mathbf{k}_{i}/\left\vert \mathbf{k}_{i}\right\vert $ is the
unit vector of $\mathbf{k}_{i}$. The general spatial profile given by Eq.~(%
\ref{Eq: m(r)-profile}) can be reduced to multiple magnetic states
including: (i) Helix (single-$\mathbf{q}$ state) when $m_{1,\parallel
}\approx m_{1,\perp }\neq 0$ and $m_{i,\parallel }=m_{i,\perp }=0$ ($i=2,3$%
); (ii) Stripe domain when only $m_{1,\perp }$ is nonzero; (iii) (nonchiral)
bubble lattice when $m_{i,\perp }\neq 0$ and $m_{i,\parallel }=0$ ($i=1,2,3$%
); (iv) SkX (triple-$\mathbf{q}$ state) when $m_{i,\parallel }\approx
m_{i,\perp }\neq 0$ and $m_{i,\perp }m_{i,\parallel }$\textit{\ }($i=1,2,3$)%
\textit{\ }have the same sign so that the three superimposed helices exhibit
the same chirality. Other magnetic states are possible as we will mention
below.

To simplify the problem without loss of generality, we shall assume that the
magnitude of the magnetization has \textit{mirror symmetry} about the $y=0$
plane, and let $
m_{2,\perp }=m_{3,\perp }\text{, and }m_{2,\parallel }=m_{3,\parallel }\,.
$
By placing Eq.~(\ref{Eq: m(r)-profile}) in Eq.~(\ref{Eq:f_tot}) and carrying out the integration in
Eq.~(\ref{Eq:f_tot}), one can obtain the spatially averaged free energy density. The expression for this is complicated and not very instructive, and is given in the supplementary material.

The global minimum of the magnetic free energy density can be computed with the following six variational parameters: $%
m_{0}, m_{1,\parallel }, m_{2,\parallel }$, $m_{1,\perp }$, $m_{2,\perp }$ and $%
q$. To specify the two GL
parameters $t$ and $u$, we introduce an extra positive definite free energy density term that arises when  the
magnetization density deviates from its uniform bulk saturation value $M_s$~\cite{jWang13IJSS_GL-penalty-func,Landis08JMPS_GL-penalty-func}, i.e., $
\delta f=\chi \left( \mathbf{m}^{2}-\mathbf{m}_{s}^{2}\right) ^{2}\,.
$
Here, $\chi$ is a positive definite parameter which in principle relies on the magnetic property of the material, and the reduced saturation magnetization $m_s(\equiv M_s/M_0)$ can be
determined by self-consistently solving the equation $M_{s}=M_{0}B_{S}\left(
T,H_{z}\right) $ in the mean field approximation with $B_{S}\left(
T,H_{z}\right) $ the Brillouin function. We thus identify $t=-2\chi
m_{s}^{2}$ and $u=\chi$. Just below $%
T_{c}$, the mean field equation gives $m_{s}\sim (1-\frac{T}{T_{c}})^{1/2}$,
by which one recovers $%
t=k_B\left( T-T_{c}\right)/2v_0 $ in the case of a 1-D Ising ferromagnet~\cite{Garel82PRB_Ginzberg-Landau} when $\chi =k_B T_{c}/4v_0$. We also impose the constraint that the spatially averaged
magnetization modulus be no greater than the saturation magnetization,
i.e., $S^{-1}\int d^{2}\mathbf{r}\left\vert \mathbf{m}\right\vert ^{2}\leq
m_{s}^{2}$.

\section{Results and discussion}
\begin{figure*}[thp]
\centering
\subfigure[~$K_u=K_{u0}$ \& $K_d=K_{d0}$] {
\includegraphics[trim={0.0cm 0.2cm 7.0cm 1.5cm},clip=true, width=0.32\linewidth]{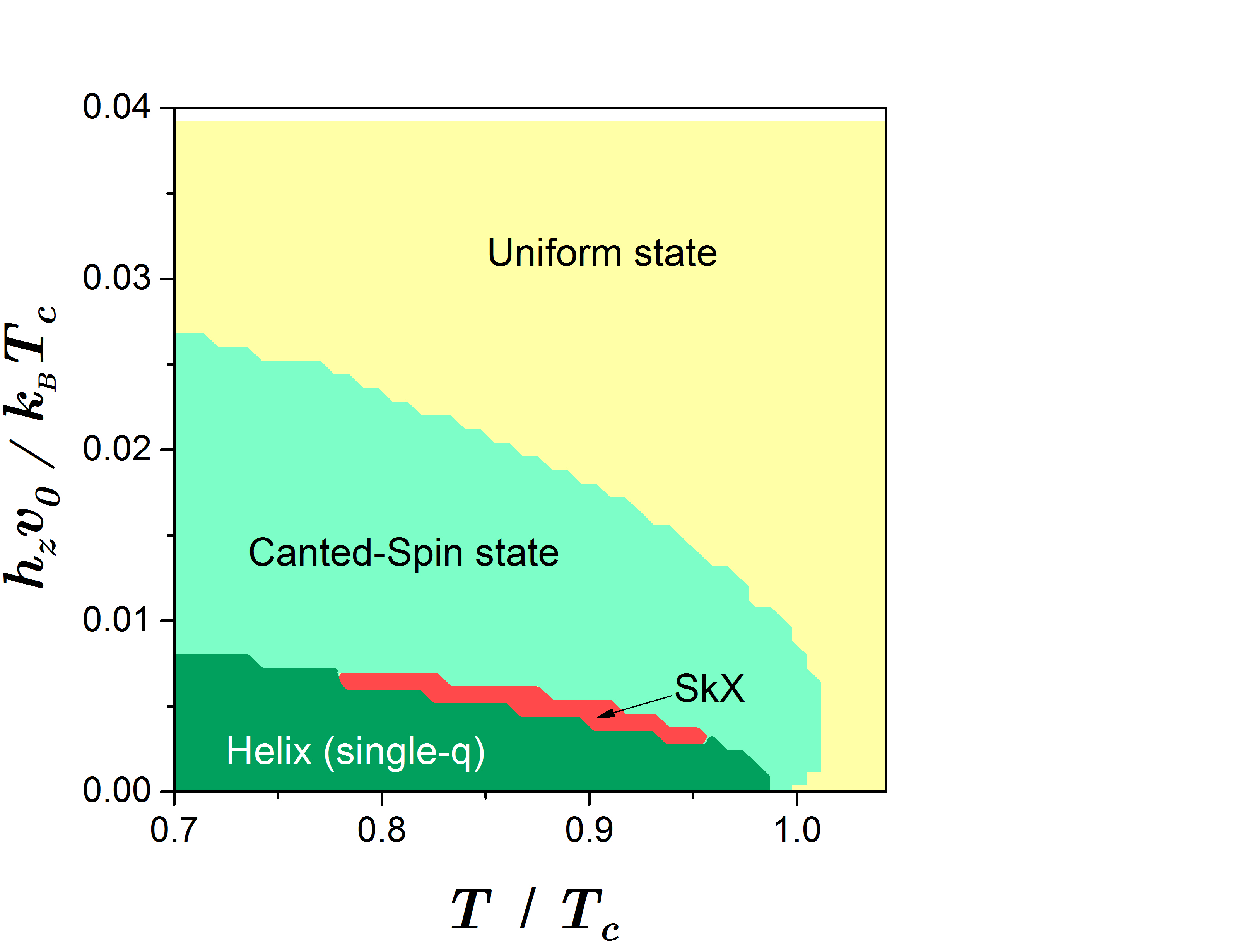}}
\subfigure[~$K_u=K_{u0}$ \& $K_d=1.5~K_{d0}$] {
\includegraphics[trim={0.0cm 0.5cm 7.0cm 1.5cm},clip=true, width=0.32\linewidth]{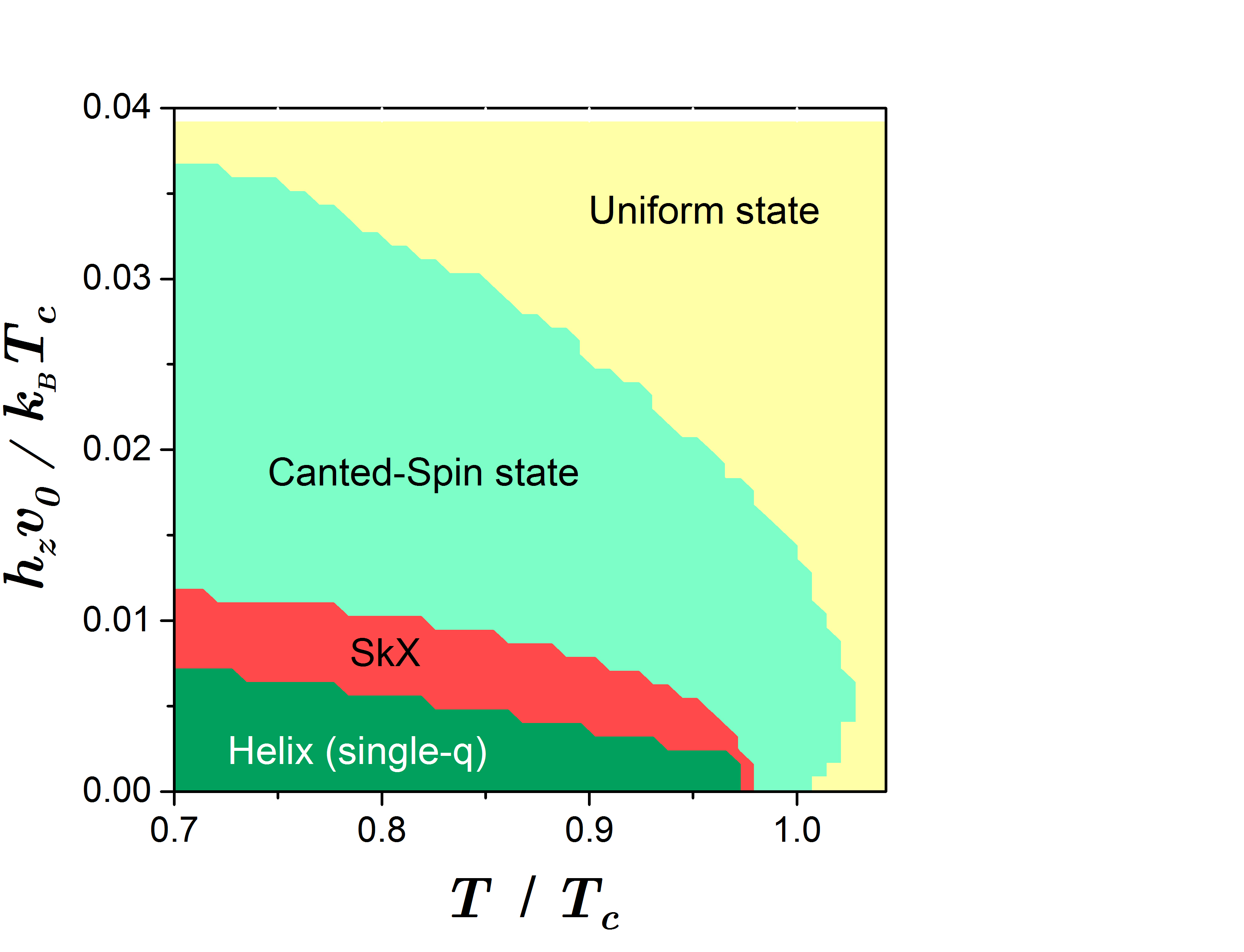}}
\subfigure[~$K_u=1.5~K_{u0}$ \& $K_d=K_{d0}$] {
\includegraphics[trim={0.0cm 0.2cm 7.0cm 1.5cm},clip=true, width=0.32\linewidth]{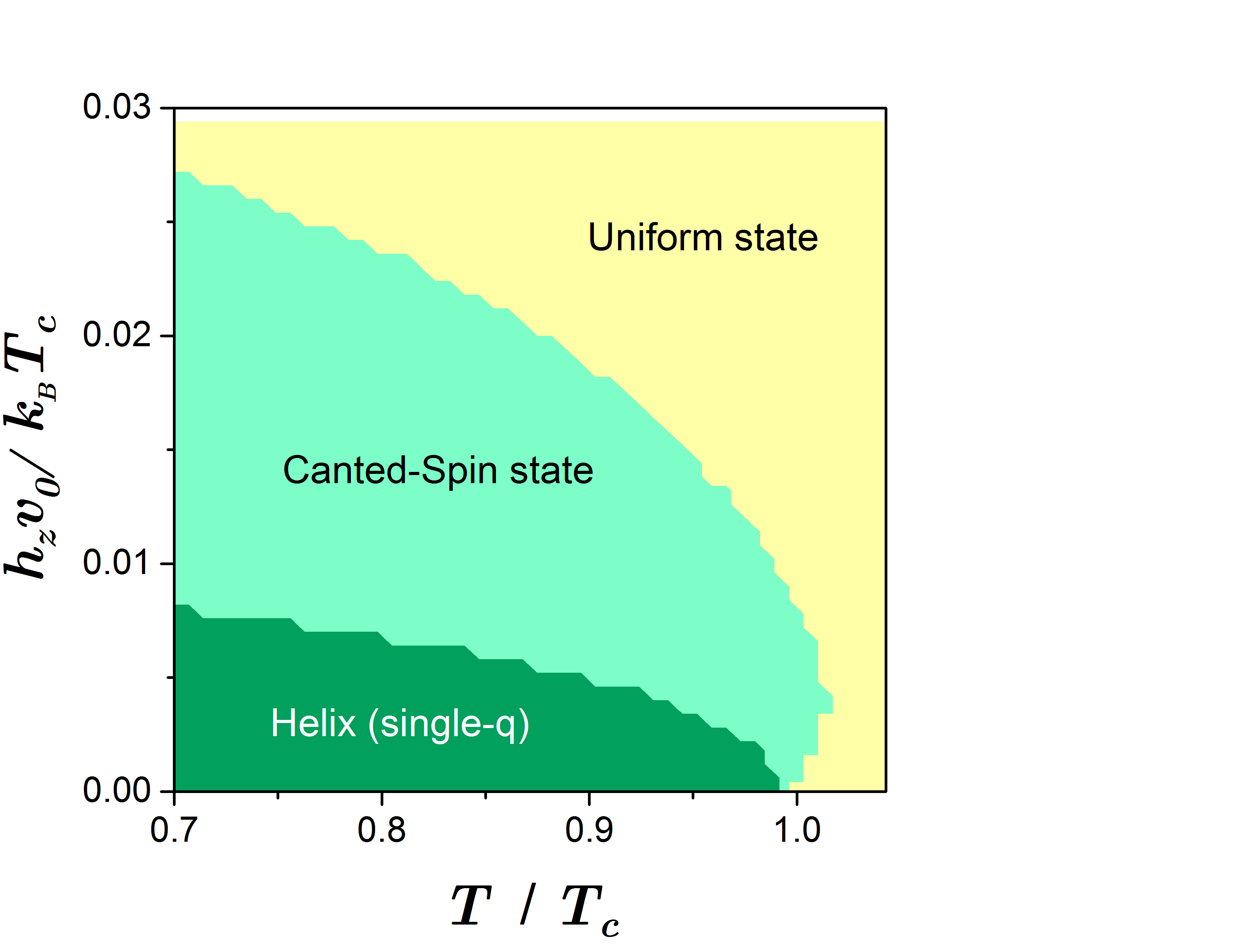}}
\caption{Phase diagram in the plane of the out-of-plane magnetic field $h_{z}(\equiv M_{0}H_{z})$ and
temperature $T$. Material parameters used in the calculation (corresponding to Ni$_2$MnGa~\cite{TICKLE99JMMM_NMG,HECZKO01JMMM_NMG,ppWu11PhilosMag_NMG,Sakon13metals_NMG}): $A_{ex}=1.0 \times 10^{-11}$~$J/m$, $M_0=6.0 \times 10^5$~A/m, $K_{u0}=2.5\times 10^5$~J/m$^3$ and $w_t=50$~nm. Small variations in the parameters may change the phase boundary, but the topology of the phase diagrams remains the same.}
\label{fig:H-T}
\end{figure*}

In Fig.~\ref{fig:H-T}, we show the phase diagram in the temperature-field plane, where the field is applied out-of-plane,
for several different magnitude
of the demagnetizing and anisotropy energy density coefficients, i.e., $%
K_{d} $ and $K_{u}$. We note that in the absence of uniaxial anisotropy,
only the uninteresting uniform magnetized state is observed (not shown).
Once the anisotropy is turned on, the helix state prevails at low magnetic
fields because of the alternating in-plane and out-of-plane easy axis variants.
When the out-of-plane magnetic field $H_{z}$ increases, the negative $z$-component
of the magnetization diminishes in a manner such that magnetizations with
an out-of-plane component start to swirl around in order to lower the exchange
energy, similar to the case in chiral magnets~\cite{Nagaosa13Nat.Nano_Skyrmion}.  This gives birth to the
SkX phase as seen in Figs~\ref{fig:H-T}~(a) and (b). We also note that, for
a given magnitude of the anisotropy, the SkX phase region is more extended
for larger $K_{d}$ since then a greater magnetic field is needed to overcome the
demagnetizing field. Further increasing the magnetic field leads to the
canted-spin state with a large out-of-plane magnetization component parallel
to the magnetic field together with small in-plane magnetization components forming an
antiferromagnetic arrangement along the $x$-axis. Also, for a given magnitude of the magnetistatic energy $%
K_{d}$, a larger anisotropy lowers the energy barrier between the helix and
canted-spin states and hence makes the SkX phase unstable, as indicated in
Figs.~\ref{fig:H-T}~(a) and (c).

\begin{figure}[tph]
\centering
\includegraphics[trim={1.8cm 1.0cm 3.0cm 2.0cm},clip=true,
width=0.45\textwidth]{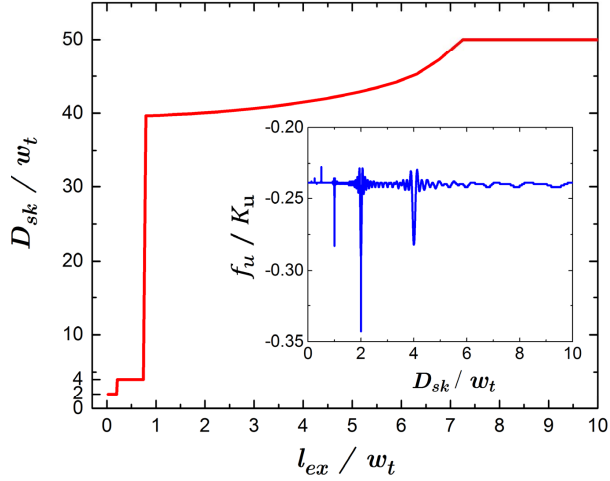}
\caption{Skyrmion diameter as a function of exchange length for a thin film
with total of 50 twin variants, i.e., $N_t=50$ and fixed width of twin width of
$w_t=50~nm$. The inset shows the coefficient of anisotropy energy density $f_u$
as a function of the diameter of a skyrmion $D_{sk}(=\frac{2\protect\pi}{q})$%
.}
\label{fig:Dsk}
\end{figure}

Next, we show that the the size of the skyrmions depends strongly on the
magnitude of the anisotropy as well as on the width of the twin variant. In
order to see this, let us focus on the free energy density of a standard
SkX state given by
\begin{align}
\bar{f}_{sk}\left( q\right) =& \left( t+K_{d}-\pi ^{-1}K_{u}\right)
m_{0}^{2}+um_{0}^{4}-h_{z}m_{0}  \notag \\
+& \frac{1}{4\pi }\left[ f_{ex}\left( q\right) +f_{u}\left( q\right)
+48um_{0}^{2}+6\pi K_{d}\right] m_{sk}^{2}  \notag \\
& +9um_{0}m_{sk}^{3}+\frac{51}{4}um_{sk}^{4}\,,
\end{align}%
where we have set $m_{i,\parallel }=m_{i,\perp }$ $=m_{sk}$ ($i=1,2,3$) in Eq.~(\ref{Eq: m(r)-profile}).
The $q$-dependence of the free energy enters through the coefficients of the exchange and anisotropy energy densities given by $
f_{ex}\left( q\right) =12\pi A_{ex}q^{2}
$ and
\begin{align}
f_{u}\left( q\right)  =-K_{u}&\left[ 9-2\zeta _{N_{t}}\left( 2w_{t}q\right)
+2\eta _{N_{t}}\left( 2w_{t}q\right) \right.   \notag \\
&\left. -\zeta _{N_{t}}\left( w_{t}q\right) +4\eta _{N_{t}}\left(
w_{t}q\right) \right] \,, \label{Eq:f_u}
\end{align}%
respectively, where $\eta _{N_{t}}\left( x\right) \equiv \frac{\sin \left[ \left(
N_{t}-1\right) x\right] +\sin \left( x\right) }{4N_{t}\left[ 1-(\frac{x}{\pi})^{2}\right] \sin \left( \frac{x}{2}\right) }$ and $\zeta _{N_{t}}\left( x\right) \equiv \frac{\sin \left[
 N_{t}x\right] }{4N_{t}\left[ 1-(\frac{x}{\pi})^{2}\right] \sin \left(
 \frac{x}{2}\right) }$ with $N_{t}$ the total number of twin variants which is taken to be an even number without loss of generality [see the supplementary material for a detailed derivation of $f_{u}(q)$].

The diameter of a skyrmion can thus be determined via $D_{sk}=\frac{2\pi }{%
q_{m}}$, where $q_{m}$ is the wave vector obtained by minimizing $%
f_{q}=f_{ex}+f_{u}$. In Fig.~\ref{fig:Dsk}, we show $D_{sk}$ as a function
of the exchange length $l_{ex}=2\pi \sqrt{\frac{A_{ex}}{K_{u}}}$ which is
the length scale of a $360^{\circ }$ Bloch domain. When $l_{ex}$ is greater
than the width of the twin $w_{t}$ (corresponding to small anisotropy $K_{u}$
for fixed exchange stiffness), the size of the skyrmion is comparable to and
eventually approaches the lateral size of the film (i.e., $L=N_{t}w_{t}=2.5$
$\mu m$ for $N_{t}=50$ and $w_{t}=50$ $nm$); in other words, the system is essentially in a uniformly magnetized state in the small anisotropy
regime. In the
intermediate anisotropy regime where $l_{ex}\lesssim w_{t}$, a plateau of $%
D_{sk}=4w_{t}$ appears, which agrees with the experimental observation of
the the close-packed hexagonal skyrmion lattice in the narrow twinned region
of Ni$_{2}$MnGa. Finally, in the large anisotropy limit where the exchange length
is much smaller than the twin width, another plateau of $D_{sk}=2w_{t}$
appears. These two plateaus result from the two local minimum in the
anisotropy density
characterized by the function $f_u$ given by Eq.~(\ref{Eq:f_u})
at $D_{sk}=2w_{t}$ and $D_{sk}=4w_{t}$, as shown by the
inset of the Fig.~\ref{fig:Dsk}.

Before we close this section, we briefly discuss the chirality and magnetic field dependence of the SkX. It was observed experimentally that the Ni$_{2}$MnGa crystal with inversion
symmetry (and thus no DMI) still may host a SkX with a single chirality, in
contrast to the SkX stabilized by long range dipolar interaction for which
the chirality of each individual skyrmion could in principle be completely random. By
adopting our {\it Ansatz} magnetization profile, we presumed all skyrmions have the same
chirality, but our analysis provides several hints about the origin of the fixed
chirality of the SkX in this system. First, as shown in the phase diagram
(Fig.~\ref{fig:H-T}), the helix state is favored at low magnetic fields. Continuous
transition from the helix state to SkX state necessitates a single chirality
of the SkX even in the absence of the DMI, since the in-plane magnetization
orientation must conform to that of the helix in order to lower the exchange
energy cost during the transition. Second, as the separation distance
between two neighboring skyrmions becomes shorter, same chirality becomes
energetically preferable considering that the in-plane magnetization
component along a line connecting the centers of the two skyrmions with
opposite chiralities would carry higher order harmonics and thus results in higher
exchange energy.

We finally comment on the dependence of the stability of the SkX hosted in the Ni$_{2}$MnGa system on the external magnetic field. Based on our theoretical model, a small but finite magnetic field is required to overcome the demagnetizing field and
stabilizing the SkX phase; the magnetic field is of the order of $50$ $Oe$ as estimated for an Ni$_2$MnGa ultrathin film of 10 monolayers at room temperature, where $h_z v_0/k_BT_c \simeq 0.05$, $v_0=a_m^2d$ with the magnetic spacing $a_m\sim 10~\mathring{A}$~\cite{RIGHI07ActaMater,Sakon13metals_NMG} and $d$ the thickness. Experimentally, SkX was observed even in the absence of external magnetic field~\cite{Phatak16NanoLett_skyrm-multiferro}. The observed (metastable) zero-field SkX may arise from the history-dependence of the system, e.g., a quenched SkX. Furthermore, we note that Ni$_{2}$MnGa
is a ferromagnetic shape memory alloy, and there is a strain energy and an associated internal
magnetic field induced by the magneto-elastic coupling involved in the martensite transformation; this is not
considered explicitly in our present model, but would be interesting for
future studies, as stabilization of skyrmions with zero magnetic field will be beneficial
for the application of skyrmions in future electronic devices.

\section{Summary and outlook}
In this work, we generalized the Ginzburg-Landau theory for Ising ferromagnets to include a general continuum magnetization profile that encapsulates various magnetic configurations with three dimensional magnetization directions. We demonstrated that stabilization of room temperature SkX can be facilitated by geometric modulation of the uniaxial anisotropy easy axis in nonchiral materials with inversion symmetry. Remarkably, the size of the skyrmions can be tailored by the period of the spatial modulation of the anisotropy, in contrast with skyrmions originating from dipolar interaction and DMI. Such novel uniaxial anisotropy was realized experimentally in a multiferroic material Ni$_2$MnGa with narrow twin variants for which the anisotropy easy axis is rotated by $90^{\circ}$ across the twin boundary, and our work explains the underlying physics that gives rise to the observed SkX in Ni$_2$MnGa.

Our model can be further generalized to take into account other magnetic interactions such as the DMI as well as various forms of geometric confinement on magnetic structures. It will also be very intriguing to investigate the transport and dynamic behaviors of skyrmions in the presence of geometric modulation of the anisotropy. For example, with the nonuniform anisotropy that we considered here, one would expect the skyrmions to respond rather differently when they are driven along and perpendicular to the twin boundaries. It may also be possible to control and alter the chirality of the SkX using, e.g., strain or electrical currents. These centrosymmetric SkXs may therefore enable interesting applications in spintronics.

\section*{Acknowledgements}
Work by S.S.-L.Z, C.P, A.P.-L, O.H was supported by Department of Energy, Office of Science, Materials Sciences and Engineering Division. Initial work by S.S.-L.Z was also partly supported by NSF Grants DMR-1406568.
\begin{figure}[h!]
\centering
\includegraphics[trim={1.8cm 21cm 1.0cm 3cm},clip=true,
width=0.5\textwidth]{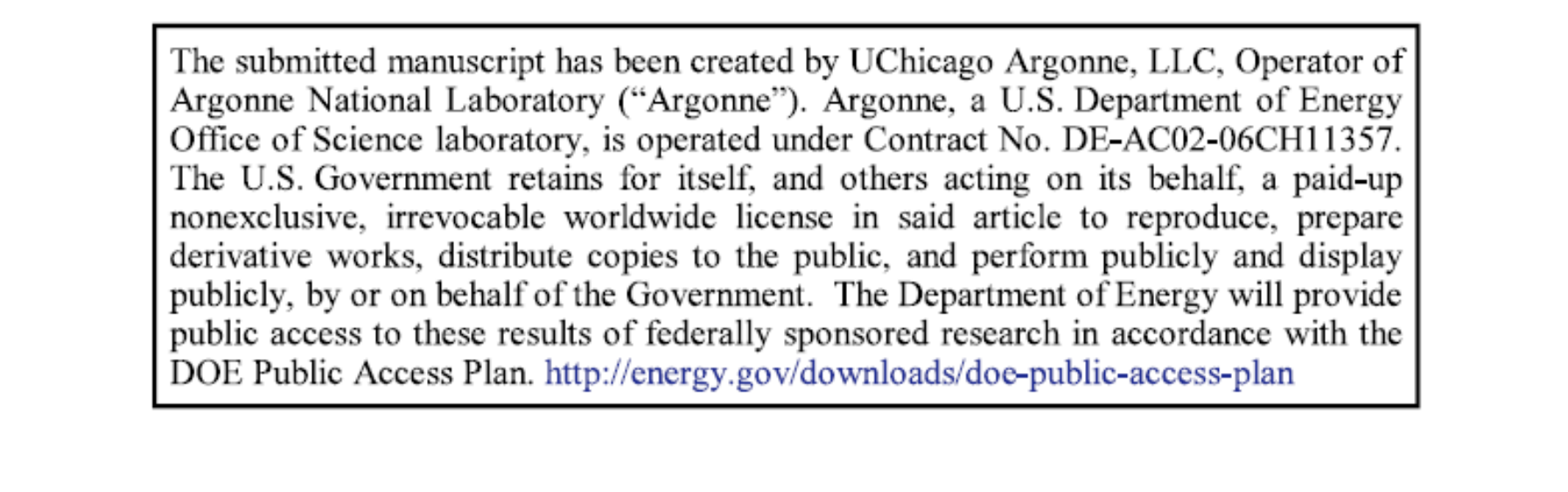}
\end{figure}

\appendix

\section{General expression of the spatially averaged free energy density}
By placing the general magnetization profile (Eq.~(3) in the main text) in the magnetic free energy expression given by Eq.~(1) in the main text, we obtain the following spatially average free energy density
\begin{align}
\bar{f}& =\left( t+K_{d}-\pi ^{-1}K_{u}\right)
m_{0}^{2}+um_{0}^{4}-h_{z}m_{0}  \notag \\
& +\frac{1}{2\pi }\left\{ \pi \left( A_{ex}q^{2}+t+2um_{0}^{2}\right) -K_{u}%
\left[ 1-\zeta _{N_{t}}\left( 2w_{t}q\right) \right] \right\} m_{1,\parallel
}^{2}  \notag \\
& +\frac{1}{2\pi }\left\{ \pi \left( A_{ex}q^{2}+t+6um_{0}^{2}+K_{d}\right)
-K_{u}\left[ 1+\eta _{N_{t}}\left( 2w_{t}q\right) \right] \right\}
m_{1,\perp }^{2}  \notag \\
& +\frac{1}{4\pi }\left\{ 4\pi \left( A_{ex}q^{2}+t+2um_{0}^{2}\right) -K_{u}%
\left[ 1-\zeta _{N_{t}}\left( w_{t}q\right) \right] \right\} m_{2,\parallel
}^{2}  \notag \\
& +\frac{1}{\pi }\left\{ \pi \left( A_{ex}q^{2}+t+6um_{0}^{2}+K_{d}\right)
-K_{u}\left[ 1+\eta _{N_{t}}\left( w_{t}q\right) \right] \right\} m_{2,\perp
}^{2}  \notag \\
& +um_{0}\left[ 2m_{1,\parallel }m_{2,\parallel }m_{2,\perp }+m_{1,\perp
}\left( m_{2,\parallel }^{2}+6m_{2,\perp }^{2}\right) \right]  \notag \\
& +\frac{u}{8}\left[ 3m_{1,\parallel }^{4}+3m_{1,\perp }^{4}+2m_{1,\parallel
}^{2}m_{1,\perp }^{2}+12m_{2,\parallel }^{4}+18m_{2,\perp
}^{4}+12m_{2,\parallel }^{2}m_{2,\perp }^{2}\right.  \notag \\
& \left. +\text{ }12\left( m_{1,\parallel }^{2}m_{2,\parallel
}^{2}+2m_{1,\perp }^{2}m_{2,\perp }^{2}\right) +8\left( m_{2,\parallel
}^{2}m_{1,\perp }^{2}+m_{2,\perp }^{2}m_{1,\parallel }^{2}\right) \right] \,. \tag{A1}
\label{Eq:ave-f}
\end{align}%
where $h_{z}=M_{0}H_{z}$, and we have assumed, for simplicity without losing generality, that the
magnitude of the magnetization has \textit{mirror symmetry} about the $y=0$
plane, and let $
m_{2,\perp }=m_{3,\perp }\text{, and }m_{2,\parallel }=m_{3,\parallel }
$; the two functions $\eta _{N_{t}}\left( w_{t}q\right)$ and $\zeta _{N_{t}}\left( w_{t}q\right)$, with $N_{t}$ the total number of twin layers, are derived from the spatial integral of the anisotropy energy density. We will present the detailed derivation of the anisotropy free energy density term in the next section.

\section{Derivation of the anisotropy free energy density term}

Let us consider the uniaxial magnetic anisotropy of the form

\begin{equation}
f_{a}\left[ \mathbf{m}\left( \mathbf{x}\right) \right] =-K_{\perp }\left(
x\right) m_{z}^{2}-K_{\parallel }\left( x\right) m_{x}^{2}  \tag{A2}
\end{equation}%
with
\begin{equation}
K_{\perp }\left( x\right) =\frac{1}{2}K_{u}\left[ \cos \left( \frac{\pi x}{%
w_{t}}\right) +\left\vert \cos \left( \frac{\pi x}{w_{t}}\right) \right\vert %
\right]  \tag{A3}
\end{equation}%
and
\begin{equation}
K_{\parallel }\left( x\right) =\frac{1}{2}K_{u}\left[ \left\vert \cos \left(
\frac{\pi x}{w_{t}}\right) \right\vert -\cos \left( \frac{\pi x}{w_{t}}%
\right) \right]\,,  \tag{A4}
\end{equation}%
where the prefactor $K_{u}$ measures the magnitude of the anisotropy energy
density, $w_{t}$ is the period of the spatial variation of the anisotropy
(or the width of the twin).

For the interesting case of $q>\pi /L$ with $L$ the side length of the
rectangular film, the spatially averaged anisotropy energy can be calculated
by the following piece-wise integration
\begin{equation}
\bar{f}_{a}=-\frac{\sqrt{3}q}{4\pi }\int_{-\frac{2\pi }{\sqrt{3}q}}^{\frac{%
2\pi }{\sqrt{3}q}}dy\sum_{n=0}^{N_{t}/2-1}\frac{K_{u}}{N_{t}w_{t}}\left[
\int_{\left( 2n-\frac{1}{2}\right) w_{t}}^{\left( 2n+\frac{1}{2}\right)
w_{t}}dx\cos \left( \frac{\pi x}{w_{t}}\right) m_{z}^{2}-\int_{\left( 2n+%
\frac{1}{2}\right) w_{t}}^{\left( 2n+\frac{3}{2}\right) w_{t}}dx\cos \left(
\frac{\pi x}{w_{t}}\right) m_{y}^{2}\right]\,,   \tag{A5}
\end{equation}%
where we have assumed $N_{t}$ being an large even integer number.  By
carrying out the integration, we obtain%
\begin{align}
\bar{f}_{a} =-\frac{K_{u}}{4\pi N_{t}}\sum_{n=0}^{N_{t}/2-1} &\left\{
2\left( 4m_{0}^{2}+2m_{1,\perp }^{2}+4m_{2,\perp }^{2}+2m_{1,\parallel
}^{2}+m_{2,\parallel }^{2}\right)+ \frac{2m_{1,\parallel }^{2}b_{n}\left( 2w_{t}q\right) }{\left(
\frac{2w_{t}q}{\pi }\right) ^{2}-1}-\frac{m_{2,\parallel }^{2}b_{n}\left(
w_{t}q\right) }{1-\left( \frac{w_{t}q}{\pi }\right) ^{2}} \right.   \notag \\
& \left. -2\left[\frac{\left( 4m_{0}m_{1,\perp }+2m_{2,\perp }^{2}\right) a_{n}\left( w_{t}q\right)}{\left( \frac{w_{t}q}{\pi }\right) ^{2}-1}+\frac{m_{1,\perp
}^{2}a_{n}\left( 2w_{t}q\right) }{\left( \frac{2w_{t}q}{\pi }\right) ^{2}-1}
\right] \right\} \,,  \tag{A6}
 \label{APP-Eq: f_a}
\end{align}%
where $a_{n}\left( x\right) =\cos \left( 2nx\right) \cos \left( \frac{x}{2}%
\right) $ and $b_{n}\left( x\right) =\cos \left[ \left( 2n+1\right) x\right]
\cos \left( \frac{x}{2}\right) $, and the results of the summation of
corresponding series are given as follows
\begin{equation}
\sum_{n=0}^{l}a_{n}\left( x\right) =\cos \left( \frac{x}{2}\right) \csc
\left( w_{t}q\right) \cos \left( lw_{t}q\right) \sin \left[ \left(
l+1\right) w_{t}q\right]  \tag{A7}   \label{APP-Eq: a_n}
\end{equation}%
and
\begin{equation}
\sum_{n=0}^{l}b_{n}\left( x\right) =\sum_{n=0}^{l}\cos \left[ \left(
2n+1\right) x\right] \cos \left( x/2\right) =\cos \left( \frac{x}{2}\right)
\csc \left( x\right) \cos \left[ \left( l+1\right) x\right] \sin \left[
\left( l+1\right) x\right]\,. \tag{A8}  \label{APP-Eq:b_n}
\end{equation}%
Placing Eqs.~(\ref{APP-Eq: a_n}) and (\ref{APP-Eq:b_n}) in Eq.~(\ref{APP-Eq:
f_a}), we obtain
\begin{align}
\bar{f}_{a} &=-\frac{K_{u}}{4\pi }\left\{ 4m_{0}^{2}+2m_{1,\perp }^{2}\left[
1+\eta _{N_{t}}\left( 2w_{t}q\right) \right] +4m_{2,\perp }^{2}\left[ 1+\eta
_{N_{t}}\left( w_{t}q\right) \right] \right.   \notag \\
&\left. +2m_{1,\parallel }^{2}\left[ 1-\zeta _{N_{t}}\left( 2w_{t}q\right) %
\right] +m_{2,\parallel }^{2}\left[ 1-\zeta _{N_{t}}\left( w_{t}q\right) %
\right] -8m_{0}m_{1,\perp }\zeta _{N_{t}}\left( w_{t}q\right) \right\}\,,  \tag{A9}
\label{APP-Eq: f_a-wx}
\end{align}%
where
\begin{equation}
\zeta _{N_{t}}\left( w_{t}q\right) =\frac{\pi ^{2}\sin \left( N_{t}w_{t}q%
\right) }{4N_{t}\left( \pi ^{2}-q^{2}w_{t}^{2}\right) \sin \left( \frac{%
w_{t}q}{2}\right) }    \tag{A10}
\end{equation}%
and
\begin{equation}
\eta _{N_{t}}\left( w_{t}q\right) =\pi ^{2}\frac{\sin \left[ \left(
N_{t}-1\right) w_{t}q\right] +\sin \left( w_{t}q\right) }{4N_{t}\left( \pi
^{2}-q^{2}w_{t}^{2}\right) \sin \left( \frac{w_{t}q}{2}\right) }\,.   \tag{A11}
\end{equation}%
Note that $\left\vert \zeta _{N_{t}}\left( x\right) \right\vert \leq \frac{1}{2}$
and $\left\vert \eta _{N_{t}}\left( x\right) \right\vert \leq \frac{1}{2}$, and  the cross-term of $m_{0}m_{1,\perp }$ in Eq.~(\ref{APP-Eq: f_a-wx}%
) vanishes when one carries out the spatial integration with an overall
sinusoidal envelop function with the period of the film length on top of the
magnetization profile [corresponding to several replica of the entire thin
film]. Taking this into account, we arrive at the final expression of the
anisotropy energy density in the presence of the geometric confinement with
only quadratic terms, i.e.,
\begin{align}
\bar{f}_{a} &=-\frac{K_{u}}{4\pi }\left\{ 4m_{0}^{2}+2m_{1,\perp }^{2}\left[
1+\eta _{N_{t}}\left( 2w_{t}q\right) \right] +4m_{2,\perp }^{2}\left[ 1+\eta
_{N_{t}}\left( w_{t}q\right) \right] \right.   \notag \\
&\left. +2m_{1,\parallel }^{2}\left[ 1-\zeta _{N_{t}}\left( 2w_{t}q\right) %
\right] +m_{2,\parallel }^{2}\left[ 1-\zeta _{N_{t}}\left( w_{t}q\right) %
\right] \right\}\,.  \tag{A12}
\end{align}
The above expression is valid when %
$q>\pi /L_{i}$, where $L_{i}$ ($i=x$ or $y$) are side lengths of the rectangular thin film.

\bibliographystyle{apsrev4-1}
\bibliography{170809_tailoring-SkX}

\end{document}